\documentclass[final]{aipproc}
\usepackage{amssymb}
\usepackage{amsmath}

\layoutstyle{6x9}

\begin{document}

\title{Radiation in Lorentz violating electrodynamics}
\author{R. Montemayor}{
  address={Instituto Balseiro and CAB,
Universidad Nacional de Cuyo and CNEA , 8400 Bariloche,
Argentina} }
\author{L.F. Urrutia}{
  address={Departamento de F\'\i sica de Altas Energ\'\i as,
Instituto de Ciencias Nucleares , Universidad Nacional Aut\'onoma
de M\'exico, Apartado Postal 70-543, 04510 M\'exico D.F.} }
\begin{abstract}
Synchrotron radiation is analyzed in the classical effective
Lorentz invariance violating model of Myers-Pospelov. Within the
full far-field approximation we compute the electric and magnetic
fields, the angular distribution of the power spectrum and the
total emitted power in the m-th harmonic, as well as the
polarization. We find the appearance of rather unexpected and
large amplifying factors, which go together with the otherwise
negligible naive expansion parameter. This opens up the
possibility of further exploring Lorentz invariance violations by
synchrotron radiation measurements in astrophysical sources where
these amplifying factors are important.
\end{abstract}
\maketitle

\section{Introduction}

Possible manifestations of quantum gravity could { arise as}
modified energy-dependent dispersion relations violating the
Lorentz symmetry. Such modifications were originally proposed in
\cite{EllisNat}, opening up the door to quantum gravity
phenomenology. The first { heuristical derivation} of such a type
of dispersion relations from a theoretical {framework}
was done in the context of { a} Loop Quantum Gravity (LQG) {
inspired model of corrections to  flat space dynamics}, leading to
a consistent extension of Maxwell electrodynamics with linear
correction terms in the quantum gravity scale \cite{GPED}. An
alternative {approximation}
to LQG, {inspired }
on Thiemann's regularization \cite{THIEM1}, extended such results
to photons\cite{AMU1} and fermions\cite{AMU2}. For a review see
for example \cite{REVURRU}. { Such estimations have to be
understood as preliminary steps in the understanding of the
semiclassical limit of LQG, which  still remains an open problem.
} An alternative proposal for a quantum gravity scale modified
Maxwell and fermionic {dynamics},
based on string theory, was developed in \cite{ELLIS2}. Recently,
an effective field theory approach to the formulation of quantum
gravity induced effects has been put forward \cite{MYPOS}. The
Gambini-Pullin (GP) \cite{GPED} and Myers-Pospelov
(MP)\cite{MYPOS} { models of} modified electrodynamics lead to
a frequency and helicity dependent speed of light. Even though
both models differ in { the} details, the radiation regime
behavior is similar, up to numerical factors of order one.

Using the synchrotron radiation from the Crab Nebulae, Jacobson et
al. have found stringent bounds upon the parameters describing the
modified dispersion relations\cite{JACOBSON}. They are based on
the synchrotron radiation cut-off frequency $\omega_{c}$, and use
a heuristic kinematical estimation of $\omega_{c}$, where photon
and electron velocities are determined by the modified dispersion
relations. These kinematical estimations do not embody
birefringence effects such as the ones appearing in the GP and MP
theories, which lead to new effects in the radiation spectrum.

Here we examine this problem from the perspective of the MP
effective theory, a dimension five modified electrodynamics which
describes the dynamics arising from linear corrections at the
quantum gravity scale. We develop a complete calculation of the
synchrotron radiation. This allows the use of additional
observational information, such as the polarization  of the
radiation, in order to set new bounds on the Lorentz violation
parameters. Some of the results have been already presented in
Ref.\cite{MU1}. In this work we restrict ourselves to the
classical regime, in such a way that the fine-tuning problems
associated with the radiative corrections of the model are not
considered \cite{MYPOS,SUDPER,COLL}.

\section{Myers-Pospelov electrodynamics}

The MP effective field theory includes Lorentz violations
generated by dimension five operators, parametrized by a vector
$V^{\mu}$, and assumed to be suppressed by a factor $M_{P}^{-1}$
($M_{P}\approx10^{19}\ GeV$ is the Planck mass). The
electromagnetic field and the dynamics of the charge are affected
by these perturbations and lead to distinctive signatures. Here we
restrict ourselves to the case $V^{\mu}=(1,\mathbf{0})$, enough
to put in evidence the main consequences of these violations.

The MP action for the electromagnetic field is
\begin{equation}
S_{MP}=\int d^{4}x\left[
-\frac{1}{4}F_{\mu\nu}F^{\mu\nu}-4\pi\,j^{\mu} A_{\mu}+{\tilde
\xi}\left(  V^{\alpha}F_{\alpha\delta}\right) (V\cdot\partial
)(V_{\beta}\tilde{F}^{\beta\delta})\right], \label{mpa}
\end{equation}
where ${\tilde \xi}=\xi/M_{P}$, with $\xi$ being a dimensionless
parameter. We take the unperturbed velocity of light in vacuum,
$c$, equal to $1$. This is a gauge theory similar to Maxwellian
electrodynamics, and implies the conservation of the electrical
current $j^{\mu}=(\rho,\mathbf{j})$. The field tensor is
$F_{\mu\nu}=\partial_{\mu}A_{\nu}-\partial_{\nu}A_{\mu}$, and the
electric and magnetic fields are given by $E_{i}=F_{0i}$ and
$B_{k}=-\frac {1}{2}\epsilon_{kij}F_{ij}$, respectively. Hence
$\mathbf{E}$ and $\mathbf{B}$ satisfy
\begin{equation}
\nabla\cdot\mathbf{B}=0\mathbf{,\;\;\;\nabla\times E}+\partial_{t}
\mathbf{B}=0.
\end{equation}
In the rest frame $V^{\alpha}=(1,\mathbf{0})$, the equations of motion derived
from the action (\ref{mpa}) are
\begin{align}
\nabla\cdot\mathbf{E} &  =4\pi\rho,\\
-\partial_{t} \mathbf{E}+\nabla\times\mathbf{B}+{\tilde \xi}
\partial_{t} \left(
-\nabla\times\mathbf{E}+\partial_{t} \mathbf{B}\right)   &
=4\pi\mathbf{j}.
\end{align}
This dynamics gives place to an energy momentum tensor $T_{\mu\nu}$ with
components
\begin{align}
4\pi\, T_{\;0}^{0} &  =\frac{1}{2}(\mathbf{E}^{2}+\mathbf{B}^{2}
)-{\tilde \xi} \mathbf{E\,\cdot\,}\partial_{t} \mathbf{B},\label{ed}\\
4\pi\, \mathbf{S} & =\mathbf{E}\times\mathbf{B}-{\tilde \xi}
\mathbf{E}\times
\partial_{t} \mathbf{E},\label{pv}
\end{align}
which, outside the sources, satisfy $\partial_{t}T^{00}+\mathbf{\nabla}
\cdot\mathbf{S}=0$. These expressions for the energy-momentum
tensor and their conservation law are exact in ${\tilde \xi}$.

Alternatively, the equation of motion for $\mathbf{A}$ in the radiation gauge
is
\begin{equation}
\partial_{t} ^{2}\mathbf{A}-\nabla^{2}\mathbf{A}+2{\tilde \xi} \nabla\times\partial_{t}
^{2}\mathbf{A}=4\pi\,\left( \mathbf{j}-(\nabla\nabla^{-2}\nabla)
\mathbf{j}\right)  \equiv4\pi\,\mathbf{j}_{T},\label{ema}
\end{equation}
which in momentum space takes the form
\begin{equation}
\left( -\omega^{2}+k^{2}-2i{\tilde\xi}
\omega^{2}\,\mathbf{k}\,\times\right)
\mathbf{A(k,}\omega\mathbf{)}=4\pi\,\mathbf{j}_{T}\mathbf{(k},\omega
\mathbf{)}.\label{emce}
\end{equation}
In the circular polarization basis, where the components of a
vector field $\mathbf{A}$ are $\mathbf{A}^{\pm}= ~
\frac{1}{2}\left[  \mathbf{A}-\left(
\mathbf{\hat{k}}\cdot\mathbf{A}\right)  \mathbf{\hat{k}}\,\pm
i(\mathbf{\hat {k}}\times\mathbf{A})\right]  $ the equations
(\ref{emce}) decouple according
to
\begin{equation}
\left(  -\omega^{2}+k^{2}\pm2{\tilde \xi} \omega^{2}k\right)
\mathbf{A}^{\pm} =4\pi\mathbf{j}_{T}^{\pm}.
\end{equation}
Thus, outside the sources the potential fields $\mathbf{A}^{\pm}$
propagate with well defined energy-dependent phase velocities
$v^{-1}_{\pm}=\sqrt{1+\omega^{2} {\tilde \xi}^{2}}\pm \omega
{\tilde \xi}$. In other words, the electromagnetic field in the MP
theory propagates in vacuum as in a { standard} dispersive
birefringent medium, where the modes with circular polarization
have well defined refraction indices
\begin{equation}
n(\lambda z)=\sqrt{1+z^{2}}+\lambda z,\label{REFIND}
\end{equation}
with $z={\tilde \xi} \omega$ and $\lambda=\pm1$.

The dynamics of a classical charged particle can be obtained from the
geometrical optics limit of a scalar charged field. For this field the MP
action is
\begin{equation}
S_{MP}=\int d^{4}x\;\left[  \partial_{\mu}\varphi^{\ast}\partial^{\mu}
\varphi-\mu^{2}\varphi^{\ast}\varphi+i{\tilde
\eta}\varphi^{\ast}\left(
V\cdot\partial\right)  ^{3}\varphi\right]  ,\label{ef}
\end{equation}
and leads to the dispersion relation $\left(  p^{0}\right)
^{2}+{\tilde \eta} \left( p^{0}\right)
^{3}=\mathbf{p}^{2}+\mu^{2}$, in the reference frame where
$V^{\alpha}=\left( 1,\mathbf{0}\right) $. Here ${\tilde
\eta}=-\eta /M_{P}$, where $\eta$ is the dimensionless constant
employed in the parameterization of Jacobson et al.. From { the
above dispersion relations}, using the minimal interaction {
for the electromagnetic field} and the geometric optics limit, we
obtain the equation of motion for a massive { charged} particle
interacting with a magnetic field, up to second order in ${\tilde
\eta}$
\begin{equation}
\mathbf{\ddot{r}}=\frac{q}{E}\left(  1-\frac{3}{2}{\tilde
\eta} E+\frac{9}{4}
{\tilde \eta}^{2} E^{2}\right)  \left(  \mathbf{v\times B}\right),
\end{equation}
with $E$ being the energy of the particle. To simplify, we will
consider { the} motion in a plane orthogonal to $\mathbf{B}$,
which corresponds to the circular orbit
\begin{equation}
\mathbf{r(}t)= R \left(  \cos\omega_{0}t,\ \sin\omega_{0}t,\ 0\right),
\label{rt}
\end{equation}
with radius $R=\beta/\omega_{0}$ and Larmor frequency
\begin{equation}
\omega_{0}=\frac{|q|B}{E}\left(  1-\frac{3}{2}{\tilde
\eta} E+\frac{9}{4}{\tilde
\eta}^{2}E^{2}
\right)  .\label{OMEGA0}
\end{equation}
The range of energies expected in astrophysical systems satisfies
$E<\left( \mu/{\tilde \eta} \right)  ^{1/2}$, with $\mu/E<<1$ and
${\tilde \eta} E<<1$ and hence the Lorentz factor $\gamma$ can be
approximated by
\begin{equation}
\gamma=\left( 1-\beta^{2} \right)
^{-1/2}\simeq\frac{E}{\mu}\left( 1+2\frac{{\tilde
\eta} E^{3}}{\mu^{2}}\right)  ^{-1/2},
\end{equation}
where we are using the standard definition\ $\beta=|\mathbf{v|} /c$.

\section{Radiation field}

The solution for the equation of motion (\ref{emce}) in the
radiation approximation can be written  as
\begin{equation}
\mathbf{A(}\omega,\mathbf{r})
=\mathbf{A}_{+}\mathbf{(}\omega,\mathbf{r}
)+\mathbf{A}_{-}\mathbf{(}\omega,\mathbf{r})
=\frac{1}{r}\frac{2n(z)} {1+n^{2}(z)}\sum_{\lambda=\pm1}n(\lambda
z)e^{in(\lambda z)\omega r}\,
\mathbf{j}^{\lambda}\mathbf{(}\omega,\mathbf{k}_{\lambda}),\label{a+-}
\end{equation}
where $\mathbf{r}=r\mathbf{\hat{n}}, \,
\mathbf{k}_{\lambda}=\omega n(\lambda z)\mathbf{\hat{n}}$. The
fields $\mathbf{A}_{+}\mathbf{(}\omega,\mathbf{r})$ and
$\mathbf{A}_{-} \mathbf{(}\omega,\mathbf{r})$ correspond to right
and left circular polarization respectively. The electric and
magnetic fields are $\mathbf{E(}\omega,\mathbf{r})
=i\omega\mathbf{A(}\omega,r)$ and
$\mathbf{B(}\omega,\mathbf{r})=\mathbf{\nabla}\times\mathbf{A(}\omega,
\mathbf{r})$. { Let us remark } that, contrary to the standard
case, $\mathbf{E}$ and $\mathbf{B}$ are not orthogonal.

Although these fields provide a complete description of the radiation, the
angular distribution of the power spectrum itself is very relevant from the
phenomenological point of view. The energy emitted can be computed using the
Poynting vector (\ref{pv})
\begin{equation}
E=\int_{-\infty}^{\infty}dt\;\mathbf{n\cdot S(}t,\mathbf{r)\;}r^{2}d\Omega,
\end{equation}
which is related to the power spectrum distribution by
\begin{equation}
E=\int_{0}^{\infty}d\omega\int d\Omega\;\frac{d^{2}E}{d\Omega d\omega}
=\int_{0}^{\infty}d\omega\int d\Omega\int dT\frac{d^{2}P(T)}{dwd\Omega
}.\label{ep}
\end{equation}
The final expression for the angular distribution of the energy
spectrum in terms of the potentials  results
\begin{equation}
\frac{d^{2}E}{d\Omega d\omega}=\frac{r^{2}\omega^{2}}{8\pi^{2}}\frac
{1+n^{2}(z)}{n(z)}\left[  \mathbf{A}_{-}(-\omega,\mathbf{r})\cdot
\mathbf{A}_{+}(\omega,\mathbf{r})+\mathbf{A}_{-}(\omega,\mathbf{r}
)\cdot\mathbf{A}_{+}(-\omega,\mathbf{r})\right]  .
\end{equation}
To compute this distribution it is only necessary to express the
products $\mathbf{A}_{\mp}(-\omega,\mathbf{r})\cdot ~
\mathbf{A}_{\pm}(\omega,\mathbf{r} )$\ in terms of the current
$\mathbf{j}(\omega,\;\mathbf{k})$ via the relation (\ref{a+-}).
This yields
\begin{equation}
\frac{d^{2}P(T)}{d\omega d\Omega}=\frac{\omega^{2}}{2\pi^{2}}\sum_{\lambda
=\pm1}\frac{n^{3}(\lambda z)}{1+n^{2}(\lambda z)}\int_{-\infty}^{\infty}d\tau
e^{-i\omega\tau}e^{i\omega n(\lambda z)\mathbf{\hat{n}}\cdot\left(
\mathbf{r(}T+\tau/2\mathbf{)-r(}T-\tau/2\mathbf{)}\right)  }\mathfrak{J}
,\label{poo}
\end{equation}
for the angular distribution of the radiated power spectrum, with
\begin{align}
\mathfrak{J}  & =\frac{1}{2}\left[  \mathbf{j}^{\ast}\left(
T+\tau/2\right) \cdot\mathbf{j}\left(  T-\tau/2\right)
-n^{-2}(\lambda z)\rho^{\ast}\left(
T+\tau/2\right)  \rho\left(  T-\tau/2\right)  \right. \nonumber\\
& \left.  -i\lambda\mathbf{\hat{n}\cdot j}^{\ast}\left(
T+\tau/2\right) \times\mathbf{j}\left(  T-\tau/2\right)  \right]
.\label{v1}
\end{align}

Another relevant characteristic of the radiation is its
polarization. To describe the polarization, we will use the
reduced Stokes parameters $\nu, q, u$, according to the
definitions of Ref. \cite{RL}, where they are written in terms of
the circular polarization basis $\mathbf{e}_{\pm}$ { and }
satisfy the constraint $1=v^{2}+q^{2}+u^{2}$. Purely circular and
linear polarizations are described by $v=\pm1$, $q=u=0$ and
$q=\pm1$, $v=u=0$, respectively.

\section{Synchrotron radiation}

In the case of { the} synchrotron radiation produced by a charge
moving in a magnetic field we have
$\rho(t,\mathbf{r})=q\delta^{3}(\mathbf{r}-\mathbf{r}(t))$ and
$\mathbf{j}(t,\mathbf{r})=\rho(t,\mathbf{r})\mathbf{v}(t)$. To be
explicit we introduce the direction of observation
\begin{equation}
\mathbf{\hat{n}}=(\sin\theta,0,\cos\theta),\label{cs}
\end{equation}
in the same coordinate system where $\mathbf{r}(t)$ was defined in Eq.
(\ref{rt}). Recalling the corresponding expressions for $\mathbf{v}(t)$ we
get
\begin{eqnarray}
\mathbf{v}(T+\tau/2)\cdot\mathbf{v}(T-\tau/2)&=&\beta^{2}\cos\omega_{0}\tau,\nonumber
\\
\mathbf{n\cdot v}(T+\tau /2)\times \mathbf{v}(T-\tau /2)&=&-\beta
^{2}\cos \theta \sin \omega _{0}\tau
\end{eqnarray}
and substituting in Eq. (\ref{v1}) we are left with
\begin{equation}
\mathfrak{J}=\frac{q^{2}}{2}\left[  \beta^{2}\cos\omega_{0}\tau-n^{-2}(\lambda
z)+i\lambda\beta^{2}\cos\theta\sin\omega_{0}\tau\right]  .\label{scriptj}
\end{equation}

The power spectrum can be computed from (\ref{poo}) by taking the
time average over the macroscopic time $T$. We are interested in
the synchrotron radiation of astrophysical objects where possible
Lorentz violations could be tested, such as the emissions by
supernova remnants (SNR) or gamma ray bursts (GRB). In such
systems, frequencies of the order of $\omega=m\omega_{0}\gtrsim1\
GeV$ are detected, and there is also evidence of electrons of
energies up to $10^{4}TeV$. {These range of energies}
correspond to a typical Larmor frequency $\omega
_{0}\simeq10^{-30}\;GeV$, which means $m\gtrsim10^{30}$, and to a
Lorentz factor $\gamma\sim10^{9}$. According to this, the
interesting regime is characterized by $m\gg1$ and $\gamma\gg1$,
but such that $m/\gamma\gg1$. In
this limit we get
\begin{equation}
\left\langle \frac{d^{2}P(T)}{d\omega d\Omega}\right\rangle =\sum
_{m=0}^{\infty}\delta({\omega}-m\omega_{0})\frac{dP_{m}}{d\Omega},
\end{equation}
with $\omega_{0}=\left(  1+\frac{{\tilde
\eta}\mu}{3}\gamma^{3}\right) |q\mathbf{B|}/\left(
\mu\gamma\right)  $ and the power angular distribution, in terms
of the Bessel functions $J_m$ (and their derivatives $J{\,}'_m$),
is
\begin{equation}
\frac{dP_{m}}{d\Omega}=\frac{\omega^{2}q^{2}}{2\pi}\left(  1+2 m
{\tilde \xi} \omega \cos\theta\right)  \left\{  \left[ J_{m}^{\,
\prime}(m\beta\sin\theta)\right] ^{2}+\left[
\frac{J_{m}(m\beta\sin\theta)}{\tan\theta}\right]  ^{2}\right\}  .
\end{equation}
This expression shows an anisotropy {to} first order in ${\tilde
\xi}$. The emission is suppressed { (enhanced)} when $m{\tilde
\xi}\omega\cos\theta\simeq -1 (+1)$. Note that both effects are
amplified by the presence of the factor $m=\omega/\omega_{0}$
which can be very large in some regimes. When $m$ is large the
radiation becomes confined to a small angular range
$\Delta\theta\simeq m^{-1/3}$ around
$\theta=\pi/2$\cite{SCHWBOOK}. Thus, if we consider the radiation
in the frontiers of the beam, i.e. $\theta\simeq\pi/2\pm
m^{-1/3}$, the anisotropy becomes significant when
$\omega\simeq\left(  \omega_{0}^{2}/{\tilde \xi}^{3}\right)
^{1/5}$. The spatially integrated power radiated in the $m^{th}$
harmonic is approximately
\begin{equation}
P_{m}  \simeq\frac{q^{2}m\omega_{0}}{\sqrt{3}\pi
R\gamma^{2}}\left[
\frac{m_{c}}{m}\kappa\left(  \frac{m}{m_{c}}\right)  -\frac{2}{\gamma^{2}
}K_{2/3}\left(  \frac{m}{m_{c}}\right)
 +2\left(  m\omega_{0}{\tilde \xi}\beta\right)  ^{2}
\frac{m^{2}}{\gamma^{2}}    K_{2/3}\left(  \frac {m}{m_{c}}\right)
\right]  ,
\end{equation}
with $m_{c}$ being the critical value $m_{c}=3\gamma^{3}/2$. For
large $m$ and $1-\beta^{2}>0$ the behavior of the power radiated
in the $m^{th}$ harmonic can be obtained using the asymptotic
expressions for the MacDonald functions $K_{m/n}$. Within this
limit there are two regimes of interest, according to the ratio
between $m$ and $m_{c}$. For the case $m/m_{c}>>1$ we get
\begin{equation}
P_{m}\simeq\frac{1}{2R}\sqrt{\frac{m}{\pi\gamma}}\left(
\frac{q^{2}B} {E}\right)  ^{2}\left(  1-3{\tilde \kappa}
E+\frac{27}{4}{\tilde \kappa}^{2}E^{2}\right) \left[ 1+2{\tilde
\xi}^{2}\left( \frac{m\omega}{\gamma}\right)  ^{2}\right]
e^{-2m/3\gamma^{3}},\label{FINRES}
\end{equation}
while the complementary range $m/m_{c}<<1$ produces
\begin{equation}
P_{m}\simeq\frac{\sqrt[3]{9}\Gamma(2/3)}{\sqrt{3}\pi}\left(
\frac{q^{2}B} {E}\right)  ^{2}\left(  1-3{\tilde \kappa}
E+\frac{27}{4}{\tilde \kappa}^{2}E^{2}\right) \left[  1+\left(
\frac{m\omega_{0}{\tilde \xi}\beta}{\sqrt[3]{2}}\right) ^{2}
\frac{m^{2}}{\gamma^{2}} \right]  m^{1/3}.\label{FINRES1}
\end{equation}

To characterize the polarization we can restrict our discussion to
$\theta =\pi/2$, { since the radiation is mostly concentrated
around the plane of the orbit}. In this case the effects of the
Lorentz violation in the current are negligible, and only the
explicit violations in the Maxwell equations affect
the polarization of the radiation. The Stokes parameters take the form
\begin{equation}
v_{m}=\frac{1-R_{m}^{2}}{1+R_{m}^{2}},\;\;\;q_{m}=\frac{2R_{m}}{1+R_{m}^{2}
}\cos(\left[  n(-z)-n(z)\right] \omega r),\;\;u_{m}=\frac{2R_{m}}{1+R_{m}^{2}
}\sin(\left[  n(-z)-n(z)\right]  \omega r),
\end{equation}
which give them  in terms of the polarization index $R_{m}$
\begin{equation}
R_{m}(\theta=\pi/2)\simeq1-2\frac{{\tilde \xi} m\omega}{\gamma^{2}}\frac{J_{m}
(m\bar{\beta})}{J_{m}^{\prime}(m\bar{\beta})},
\end{equation}
where $\bar{\beta}=\sqrt{1-\mu^{2}/E^{2}}$. In the large $m$ {
limit} and in the case $1-\beta^{2}>0$, one can write the Bessel
function and its derivative in terms of the MacDonald functions.
The latter can be further approximated according to
$m/\hat{m}_{c}\lessgtr1$ where $\hat{m}_{c}=3\left(
1-\bar{\beta}^{2}\right)  ^{-3/2}/2$. In such a way, if
$m/\hat{m}_{c}>>1$ the
radiated power is exponentially damped and we get
\begin{equation}
R_{m}(\theta=\pi/2)\simeq1-2{\tilde \xi} \omega m/\gamma.
\end{equation}
Then we have, to first order in ${\tilde \xi}$,
\begin{equation}
v_{m}\simeq2\omega m{\tilde \xi}/\gamma\ ,\ \ \ \
q_{m}\simeq\cos(2{\tilde \xi}\omega ^{2}r)\ ,\ \ \ \ \ \ \
u_{m}\simeq-\sin(2{\tilde \xi}\omega^{2}r) .
\end{equation}
In the opposite case, when $m/\hat{m}_{c}<<1$, we get
\begin{equation}
R_{m}(\theta=\pi/2)\simeq1-\frac{2{\tilde \xi}\omega}{\gamma^{2}}\left(  \frac{1}
{4}\right)  ^{1/3}\frac{\Gamma(1/3)}{\Gamma(2/3)}m^{4/3},
\end{equation}
from which we have
\begin{equation}
v_{m}\simeq\left(  \frac{1}{4}\right)
^{1/3}\frac{\Gamma(1/3)}{\Gamma
(2/3)}\frac{2\omega}{\gamma^{2}}m^{4/3}{\tilde \xi}\ ,\ \ \ \ \
q_{m}\simeq\cos(2{\tilde \xi}\omega^{2}r)\ ,\ \ \ \ \ \
u_{m}\simeq-\sin(2{\tilde \xi}\omega^{2}r).
\end{equation}
In both cases powers of $m$ appear as  amplifiying factors. Note
that $q_{m}$ and $u_{m}$ oscillate with a wave length
$\Lambda=\pi/({\tilde \xi}\omega^{2})$. These oscillations come
from the term $\pm{\tilde \xi}\omega^{2}r$ { of} the phases in Eq.
(\ref{a+-}), i.e. they are a consequence of the birefringence of
the vacuum.
\section{Final remarks}
The main results of a complete calculation of synchrotron
radiation in the MP effective electrodynamics has been presented.
This effective electrodynamics can be interpreted in terms of {
a standard
parity violating birefringent and dispersive medium}. A priori one
can expect that the Lorentz violating effects would be of order
${\tilde \xi}\omega$. Instead
we find amplifying factors
proportional to $m$, which could render these modifications within
or near the scope of present time observations.

When an astrophysical object is considered as a source of
synchrotron radiation it is necessary to take into consideration
the actual range of validity of the radiation approximation {
from }where these results have been obtained. The propagation of
the fields is characterized by the phase
\begin{equation}
n(\lambda z)\omega\left\vert
\mathbf{r}-\mathbf{r}^{\prime}\right\vert \simeq\omega r\left(
1-\frac{\mathbf{n}\cdot \mathbf{r}^{\prime}}{r}+\lambda{\tilde
\xi} \omega -\lambda{\tilde
\xi}\omega\frac{\mathbf{n}\cdot \mathbf{r}^{\prime}}{r}+\frac{1}{2}\frac{r^{\prime2}
}{r^{2}}\right),
\end{equation}
{ where $r^{\prime}$ can be estimated by the radius of the
orbit.}
In the radiation approximation the term proportional to $\left(  r
^\prime
/r\right)  ^{2}$ is neglected. If $\left\vert \xi\omega\right\vert >r
^\prime
/r$ we can neglect only the term quadratic in $r^{\prime}$and both
${\tilde \xi}$-dependent terms remain in the phase, which can be
approximated, { using }
$\mathbf{k}_{\lambda}=\omega n(\lambda z)\mathbf{\hat{n}}$, { by}
\begin{equation}
n(\lambda z)\omega\left\vert \mathbf{r}-\mathbf{r}^{\prime}\right\vert \simeq
n(\lambda z)\omega r\left(  1-\mathbf{\hat{n}}\cdot\mathbf{r}^{\prime
}/r\right).
\end{equation}
This is the full far-field approximation developed in the
preceding sections, with the most general dependence on the index
of refraction. An intermediate
situation happens when $\left(  r
^\prime{}
/r\right)  ^{2}<\left\vert {\tilde
\xi}\omega\right\vert <r
^\prime{}
/r$, in which case
\begin{equation}
n(\lambda z)\omega\left\vert \mathbf{r}-\mathbf{r}^{\prime}\right\vert \simeq
n(\lambda z)\omega r-\mathbf{\hat{n}}\cdot\mathbf{r}^{\prime},
\end{equation}
and hence the index of refraction { still} remains significant.
Finally, if $\left\vert {\tilde
\xi}\omega\right\vert <\left(  r
^\prime{}
/r\right)  ^{2}$ all the dependence on ${\tilde \xi}$ is
negligible in the phase, which
reduces to the usual expression
\begin{equation}
n(\lambda z)\omega\left\vert \mathbf{r}-\mathbf{r}^{\prime}\right\vert
\simeq\omega r-\mathbf{\hat{n}}\cdot\mathbf{r}^{\prime}.
\end{equation}

To be concrete we can consider the Crab Nebula, an SNR at $6000\
ly$. The synchrotron radiation spectrum shows a cutoff at
$\omega\simeq0.5\ GeV$, and there is evidence of electrons of
energy $E_{e}\simeq3\times10^{6}\;GeV $ producing this emission.
We can estimate the magnetic field, the $\gamma$ factor and the
Larmor frequency using the unperturbed relations which give us
$B\simeq10^{-3}\ G$, $\gamma\simeq6.0\times10^{9} $, and $\omega_{0}
\simeq2\times10^{-30}\ GeV$. Thus, in this case we have $r\sim10^{36}
\ GeV^{-1}$ and $r^{\prime}\sim10^{30}\ GeV^{-1}$. Considering
that ${\tilde \xi}\lesssim10^{-19}\ GeV^{-1}$, we get that not
only $\left\vert {\tilde \xi} \omega\right\vert <r^\prime{} /r$,
but also
$\left\vert {\tilde \xi}\omega\right\vert <\left( r^\prime
/r\right)  ^{2}$. Therefore all the information about the ${\tilde
\xi}$ parameter is erased by the radiation approximation and the
power spectrum only {contains }
the parameter ${\tilde \eta}$, in agreement with Ref.
\cite{JACOBSON}. When radiation from much more distant objects
such as GRBs is considered, information about the parameter
${\tilde \xi}$ { may become} accessible. For example, { in
the case of} Mkn 501 { we have:} $r\sim10^{8}\ l.y.\sim10^{40}\
GeV^{-1}$ and $r^{\prime}\sim 10^{26}\ GeV^{-1}$, which give
$r^{\prime}/r\sim10^{-14}$. The photon energy is
$\omega\sim10^{4}\ GeV$, and thus $\left\vert {\tilde
\xi}\omega\right\vert
\lesssim10^{-15}$. In this way we have $\left(  r^{\prime}/r\right)  ^{2}
<{\tilde \xi}\omega<r^{\prime}/r$, and hence the term ${\tilde
\xi}\omega r$ in the phase becomes significant. Another example is
GRB021206, at a distance of $10^{10}\ l.y.$, with a larger
magnetic field and therefore a smaller radius for the electron
orbits. Here $r^{\prime}/r\sim10^{-24}$, $\omega\sim10^{-3}\ GeV$,
and $\left\vert {\tilde \xi}\omega\right\vert \lesssim10^{-22}$,
which yields $r^{\prime}/r\ll\left\vert {\tilde
\xi}\omega\right\vert $. In principle this case could allow access
to the full far-field approximation. { Summarizing, in} the
case of the CRAB nebula radiation the far-field approximation
erases all the information about the ${\tilde \xi}$ parameter and
only allows to test the parameter ${\tilde \eta}$ associated to
the dynamics of the charged particle. The situation changes for
extragalactic sources, such as in the case of GRB's. For these
objects the far-field approximation could {include}
${\tilde \xi}$-LV terms and the $m$ -amplifying factors become
important, { thus }giving access to a new sector of the Lorentz
{ violating parameters}.

\bibliographystyle{aipproc}

\end{document}